\documentclass{emulateapj}

\begin{document}

\shortauthors{Gordon et al.}
\shorttitle{SAGE-SMC Overview}

\title{Surveying the Agents of Galaxy Evolution in the
Tidally-Stripped, Low Metallicity Small Magellanic Cloud
(SAGE-SMC). I. Overview}

\author{K.~D.~Gordon\altaffilmark{1}, 
   M.~Meixner\altaffilmark{1},
   M.~R.~Meade\altaffilmark{2},
   B.~Whitney\altaffilmark{2,3},
   C.~Engelbracht\altaffilmark{4},
   C.~Bot\altaffilmark{5},
   M.~L.~Boyer\altaffilmark{1},
   B.~Lawton\altaffilmark{1},
   M.~Sewi{\l}o\altaffilmark{6},
   B.~Babler\altaffilmark{2},
   J.-P.~Bernard\altaffilmark{7},
   S.~Bracker\altaffilmark{2},
   M.~Block\altaffilmark{4},
   R.~Blum\altaffilmark{8},
   A.~Bolatto\altaffilmark{9},
   A.~Bonanos\altaffilmark{10},
   J.~Harris\altaffilmark{8},
   J.~L.~Hora\altaffilmark{11},
   R.~Indebetouw\altaffilmark{13,14},
   K.~Misselt\altaffilmark{4},
   W.~Reach\altaffilmark{13},
   B.~Shiao\altaffilmark{1},
   X.~Tielens\altaffilmark{15},
   L.~Carlson\altaffilmark{6},
   E.~Churchwell\altaffilmark{2},
   G.~C.~Clayton\altaffilmark{16},
   C.-H.~R.~Chen\altaffilmark{12},
   M.~Cohen\altaffilmark{17},
   Y.~Fukui\altaffilmark{18},
   V.~Gorjian\altaffilmark{19},
   S.~Hony\altaffilmark{20},
   F.~P.~Israel\altaffilmark{15},
   A.~Kawamura\altaffilmark{18,31},
   F.~Kemper\altaffilmark{21,22},
   A.~Leroy\altaffilmark{14},
   A.~Li\altaffilmark{23},
   S.~Madden\altaffilmark{20},
   A.~R.~Marble\altaffilmark{4,24},
   I.~McDonald\altaffilmark{25},
   A.~Mizuno\altaffilmark{18},
   N.~Mizuno\altaffilmark{18},
   E.~Muller\altaffilmark{18,31},
   J.~M.~Oliveira\altaffilmark{25},
   K.~Olsen\altaffilmark{8},
   T.~Onishi\altaffilmark{18},
   R.~Paladini\altaffilmark{13},
   D.~Paradis\altaffilmark{13},
   S.~Points\altaffilmark{26},
   T.~Robitaille\altaffilmark{11},
   D.~Rubin\altaffilmark{20},
   K.~Sandstrom\altaffilmark{27},
   S.~Sato\altaffilmark{18},
   H.~Shibai\altaffilmark{18},
   J.~D.~Simon\altaffilmark{28},
   L.~J.~Smith\altaffilmark{1,29},
   S.~Srinivasan\altaffilmark{32},
   U.~Vijh\altaffilmark{30},
   S.~Van~Dyk\altaffilmark{13},
   J.~Th.~van~Loon\altaffilmark{22}, \&
   D.~Zaritsky\altaffilmark{4}}

\altaffiltext{1}{Space Telescope Science Institute, 3700 San Martin Drive, Baltimore, MD 21218}
\altaffiltext{2}{Dept.\ of Astronomy, Univ.\ of Wisconsin-Madison, 475 N. Charter St., Madison, WI 53706}
\altaffiltext{3}{Space Sci.\ Inst., 4750 Walnut St., Suite 205, Boulder, CO 80301}
\altaffiltext{4}{Steward Observatory, University of Arizona, Tucson, AZ 85721}
\altaffiltext{5}{UMR 7550, Centre de Donn\'ees Astronomique de Strasbourg (CDS), Universit\'e Louis Pasteur, 67000 Strasbourg, France}
\altaffiltext{6}{Department of Physics and Astronomy, Johns Hopkins
  University, Baltimore, MD 21218}
\altaffiltext{7}{CESR, Universit\'e de Toulouse, UPS, 9 Avenue du Colonel Roche, F-31028 Toulouse, Cedex 4, France}
\altaffiltext{8}{National Optical Astronomy Observatory, 950 North Cherry Avenue, Tucson, AZ 85719}
\altaffiltext{9}{Dept.\ of Astronomy, Univ.\ of Maryland, College Park, MD 20742}
\altaffiltext{10}{Institute of Astronomy \& Astrophysics, National Observatory of Athens, I. Metaxa \& Vas. Pavlou St., P. Penteli, 15236 Athens, Greece}
\altaffiltext{11}{Harvard-Smithsonian, CfA, 60 Garden St., MS 65, Cambridge, MA 02138}
\altaffiltext{12}{Department of Astronomy, University of Virginia,
Charlottesville, VA 22904, USA}
\altaffiltext{13}{Infrared Processing and Analysis Center, Caltech, MS 220-6, Pasadena, CA 91125}
\altaffiltext{14}{National Radio Astronomy Observatory, 520 Edgemont
Road, Charlottesville, VA 22903, USA}
\altaffiltext{15}{Sterrewacht Leiden, Leiden University, P.O. Box 9513, 2300 RA Leiden, The Netherlands}
\altaffiltext{16}{Department of Physics \& Astronomy, Louisiana State
University, Baton Rouge, LA 70803} 
\altaffiltext{17}{Monterey Institute for Research in Astronomy, Marina, CA 93933}
\altaffiltext{18}{Dept.\ of Astrophysics, Nagoya Univ., Furo-cho, Chikusa-ku, Nagoya 464-8602, Japan}
\altaffiltext{19}{Jet Propulsion Laboratory, 4800 Oak Grove Boulevard, MS 169-327, Pasadena, CA 91109}
\altaffiltext{20}{Service d'Astrophysique, CEA/Saclay, l'Orme des Merisiers, 91191 Gif-sur-Yvette, France}
\altaffiltext{21}{Academia Sinica, Institute of Astronomy and Astrophysics, P.O. Box 23-141, Taipei 10617, Taiwan, R.O.C.}
\altaffiltext{22}{Jodrell Bank Centre for Astrophysics, Alan Turing Building, University of Manchester, Oxford Road, Manchester, M13 9PL, UK}
\altaffiltext{23}{Dept.\ of Physics \& Astronomy, Univ.\ of Missouri, Columbia MO 65211}
\altaffiltext{24}{National Solar Observatory, Tucson, AZ 85719}
\altaffiltext{25}{Astrophysics Group, Lennard-Jones Laboratories, Keele University, ST5 5BG, UK}
\altaffiltext{26}{National Optical Astronomy Observatory, Cerro Tololo Inter-American Observatory, La Serena, Chile}
\altaffiltext{27}{Max-Planck-Institut f\"ur Astronomie, D-69117 Heidelberg, Germany}
\altaffiltext{28}{Observatories of the Carnegie Institution of Washington, 813 Santa Barbara Street, Pasadena, CA 91101}
\altaffiltext{29}{European Space Agency, Research and Scientific Support Department, Baltimore, MD 21218}
\altaffiltext{30}{Ritter Astrophysical Research Center, University of
  Toledo, Toledo, OH 43606}
\altaffiltext{31}{ALMA-J Project Office, National Astronomical Observatory of Japan, 2-21-1 Osawa, Mitaka, Tokyo 181-8588, Japan}
\altaffiltext{32}{Institut d'Astrophysique de Paris, 98 bis, Boulevard Arago, Paris 75014, France}

\begin{abstract} 
The Small Magellanic Cloud (SMC) provides a unique laboratory for the
study of the lifecycle of dust given its low metallicity
($\sim$1/5~solar) and relative proximity ($\sim$60~kpc).  This
motivated the SAGE-SMC (Surveying the Agents of Galaxy Evolution in
the Tidally-Stripped, Low Metallicity Small Magellanic Cloud) {\em Spitzer}
Legacy program with the specific goals of studying the amount and type
of dust in the present interstellar medium, the sources of dust in the
winds of evolved stars, and how much dust is consumed in star
formation.  This program mapped the full SMC (30~$\sq\arcdeg$)
including the Body, Wing, and Tail in 7 bands from 3.6 to 160~\micron\
using the IRAC and MIPS instruments on the {\em Spitzer Space Telescope}.
The data were reduced, mosaicked, and the point sources measured using
customized routines specific for large surveys.  We have made the
resulting mosaics and point source catalogs available to the
community.  The infrared colors of the SMC are compared to those of
other nearby galaxies and the 8~\micron/24~\micron\ ratio is somewhat lower and the
70~\micron/160~\micron\ ratio is somewhat higher than the average.  The
global infrared spectral energy distribution shows that the SMC has
approximately 1/3 the aromatic emission/PAH (polycyclic aromatic hydrocarbon)
abundance of most nearby galaxies.  Infrared color-magnitude
diagrams are given illustrating the distribution of different
asymptotic giant branch stars and the locations of young stellar
objects.  Finally, the average spectral energy distribution (SED) of
HII/star formation regions is compared to the equivalent Large
Magellanic Cloud average HII/star formation region SED.  These
preliminary results will be expanded in detail in companion papers.
\end{abstract}


\section{Introduction}
\label{sec_intro}

The interstellar medium (ISM) plays a central role in galaxy
evolution as the birthsite of new stars and repository of old 
stellar ejecta. The formation of new stars slowly consumes the ISM,
locking it up for millions to billions of years. As these stars age,
the winds from low mass, asymptotic giant branch (AGB) stars, high 
mass, red supergiants (RSGs)/Wolf Rayet stars/Luminous
 Blue Variables (LBVs), and supernova explosions inject 
nucleosynthetic products of stellar interiors into the ISM, slowly
increasing its metallicity. This constant recycling and associated
enrichment drives the evolution of a galaxy's visible matter and
changes its emission characteristics. To understand this recycling, we
have to study the physical processes of the ISM, the formation of new
stars, and the injection of mass by evolved stars, and their
relationships on a galaxy-wide scale.

Among the nearby galaxies, the \object{Small Magellanic Cloud} (SMC) represents
a unique astrophysical laboratory for studies of the lifecycle of the
ISM, because of its proximity \citep[$\sim$60
kpc,][]{Hilditch05}, low ISM metallicity \citep[1/5--1/8
Z$_{\sun}$,][]{Russell92, Rolleston99, Rolleston03, Lee05} and history of disruption 
by tidal interaction \citep{Zaritsky00}.  The SMC offers a rare glimpse
into the 
physical processes in an environment with a metallicity which is below
the threshold of 1/4--1/3 Z$_{\sun}$ where the properties of the ISM
in galaxies changes significantly as traced by the rapid reduction in
the aromatic emission/polycyclic aromatic hydrocarbon (PAH) dust mass
fractions and dust-to-gas ratios 
\citep{Engelbracht05SB, Draine07, Sandstrom10}.  In addition, the SMC is the only
local galaxy which has the ultraviolet dust characteristics \citep[lack of
2175~\AA\ extinction bump,][]{Gordon03} of starburst galaxies
in the local \citep{Calzetti94, Gordon97} and high-redshift
\citep[$2<z<4$,][]{Vijh03} universe.  The evolution of
stars in the SMC is also clearly affected by their low metallicities
\citep{Cioni06, Marigo08} with the corresponding expected differences in
stellar mass loss \citep{vanLoon08}.
The Large and Small Magellanic Clouds represent
the nearest example of tidally interacting galaxies and the Magellanic
Bridge is a clear manifestation of a close encounter of these two
galaxies some 200 Myr ago \citep{Zaritsky04, Harris07}.  Over
cosmological timescales, galaxy interactions are one of the key
drivers of galaxy evolution and, thus, tidally interacting galaxies
allow us to examine star formation in an unusual and disturbed
environment, which resembles the conditions in the early universe when
galaxies were forming.  The Magellanic Bridge is a filament of neutral
hydrogen, which joins the SMC and LMC over some 15 kpc
\citep{McGee86, Staveley-Smith98, Muller03}. Recent studies have
revealed the presence of young ($<$200 Myrs) massive stars
\citep{Harris07} associated with the highest-density portion of the
Bridge directly adjacent to the SMC main body.  Given the relative
youth of these stars, they are highly likely to have been formed in
situ making this structure a tidal tail from the SMC.

The SAGE-SMC (Surveying the Agents of Galaxy Evolution in the
Tidally-Stripped, Low Metallicity Small Magellanic Cloud) program is a
{\em Spitzer} cycle 4 Legacy (285 hours, PI: K.\ Gordon, PID: 40245) to map the full SMC
(30~$\sq\arcdeg$) including the Body, Wing, and Tail using the Infrared
Array Camera \citep[IRAC,][]{Fazio04IRAC} and Multiband Imaging
Photometer for {\em Spitzer} \citep[MIPS,][]{Rieke04MIPS} instruments on the
{\em Spitzer} Space Telescope \citep{Werner04Spitzer}.  The SAGE-SMC program
builds on the pathfinder S$^3$MC program 
\citep{Bolatto07} that surveyed the inner $\sim$3~$\sq\arcdeg$ of this
galaxy.  The SAGE-SMC observations allow us to trace the life cycle of
dust (and thereby gas) on a galaxy wide scale from their injection by
late-type stars, through their sojourn in the violent ISM, until their
involvement in the formation of stars. In addition, the infrared
(IR)
emission traces the global structure of the ISM on a galaxy-wide scale
and the interrelationship of the various phases of the ISM. This
survey provides a complete census of the massive star formation population in
this low and spatially varying metallicity environment.  With much
improved wavelength coverage, up to $\sim$1000 times better point
source sensitivity and $\sim$13 times better angular resolution than
the Midcourse Space Experiment ({\em MSX}) and Infrared Astronomical
Satellite ({\em IRAS}) surveys, SAGE-SMC significantly improves our  
understanding of this important galaxy.

Combining the results from this SMC survey with the existing
LMC \citep[SAGE-LMC,][]{Meixner06} and Milky Way \citep[GLIMPSE,
MIPSGAL,][]{Churchwell09, Carey09} surveys provides a foundation for
understanding the physics of the ISM, current star formation, and
evolved-star mass loss as a function of metallicity.  This
foundation is crucial for interpreting the observations of more
distant galaxies like those in the SINGS
\citep{Kennicutt03SINGS}, SWIRE \citep{Lonsdale03}, and GOODS
\citep{Dickinson03} {\em Spitzer} Legacy
programs.  The SAGE-SMC survey provides a crucial link in our
understanding of galaxies during low metallicity, chemically-primitive
stages.

This paper presents the overview of the SAGE-SMC Legacy program
including the science motivation, observation details, data reduction,
point source catalog creation, and publicly available data
products.  In addition, results from all three areas
(interstellar medium, star formation, and stellar mass loss) of
scientific interest of the SAGE-SMC team are presented.  

\section{Data}

\begin{figure*}[tbp]
\epsscale{1.1}
\plotone{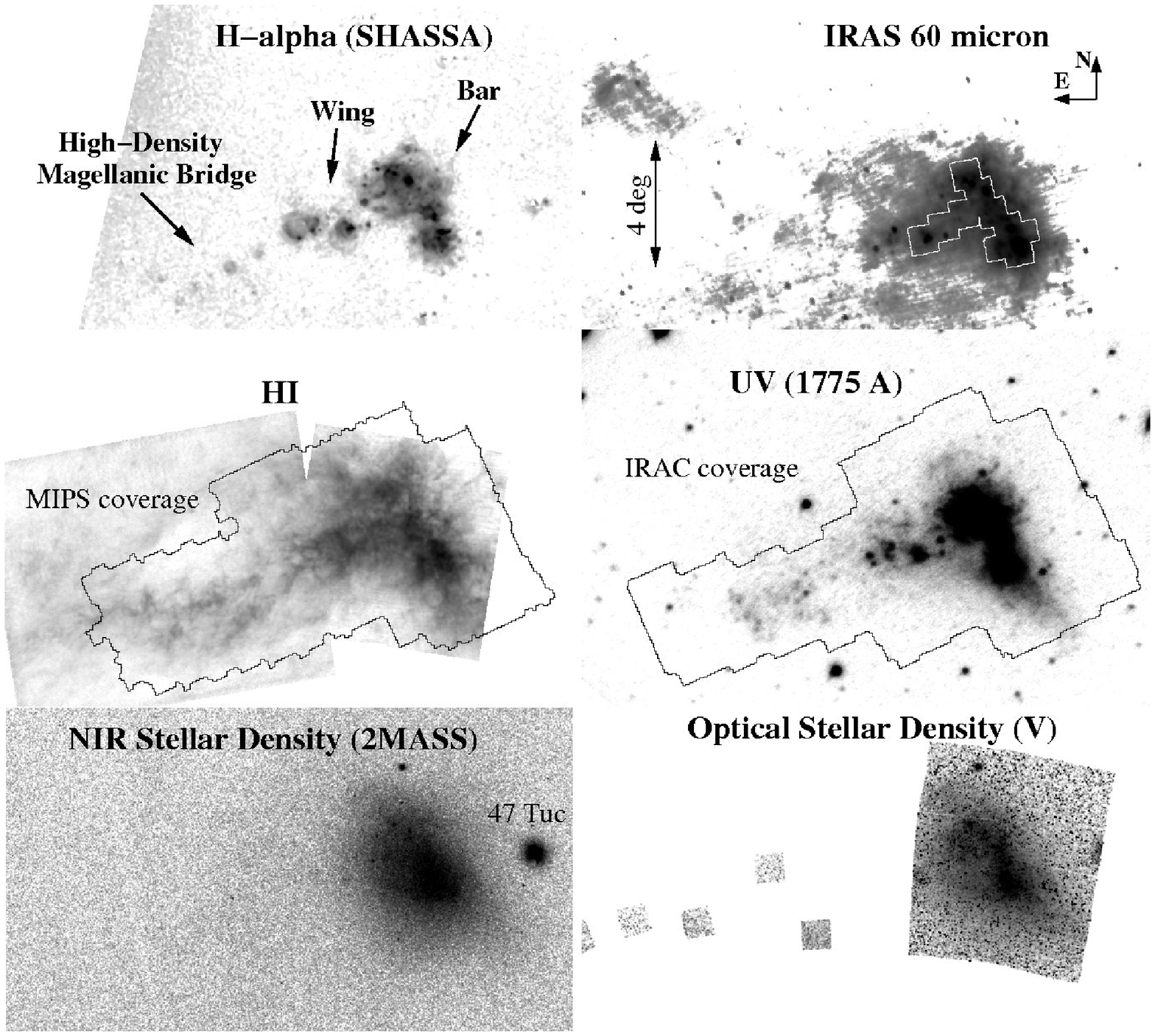}
\caption{The view of the whole SMC (Bar, Wing, and Tail) is shown in
H$\alpha$ \citep{Gaustad01}, {\em IRAS} 60~\micron\
\citep{Schwering89}, HI
\citep{Stanimirovic00, Muller03}, ultraviolet \citep[FUVCAM,][for
camera details]{Gordon94}, NIR stellar density \citep{Skrutskie06},
and optical stellar density \citep{Zaritsky00, Harris07}.  These
images are centered at (ra, dec) = (23$\fdg$09, -72$\fdg$61) and are
12$\arcdeg$ by 8$\arcdeg$ in size.  The Milky Way globular cluster 47 Tuc is
clearly seen on the NIR stellar density image.  The S$^3$MC survey
region where both IRAC and MIPS observations exist is shown
superimposed on the {\em IRAS} 60~\micron\ image.  The SAGE-SMC
coverage in MIPS and IRAC is shown overlaid on the HI and UV images,
respectively.
\label{fig_smc_mwave}}
\end{figure*}

A uniform, unbiased survey of the whole SMC ($\sim$$30~\sq\arcdeg$)
including the Bar, Wing, and Tail (Fig.~\ref{fig_smc_mwave}), in all
the IRAC (3.6, 4.5, 5.8 and 
8~\micron) and MIPS (24, 70 and 160~\micron) bands
is the basis of the SAGE-SMC Legacy
program. The angular resolutions of the different bands are 2\arcsec\
(0.6~pc, IRAC), 
6\arcsec\ (1.7~pc, MIPS 24~\micron), 18\arcsec\ (5.2~pc, MIPS 70~\micron), and
40\arcsec\ (12~pc, MIPS 160~\micron).  Previous SMC
infrared surveys have been done with {\em IRAS} at a resolution of $\sim$4$\arcmin$ for 12, 25, 60, and
100~\micron\ \citep{Schwering89, Miville-Deschenes05}, with the
Infrared Space Observatory ({\em ISO}) at a resolution of 2$\arcmin$ at
170~\micron\ \citep{Wilke03}, and with {\em MSX} at a resolution of 18$\arcsec$ at 8~\micron\ \citep{Price01}.
SAGE-SMC provides an improvement in resolution that ranges from a
factor of 9 at 8~\micron\ to 13 at 70~\micron\ to 3 at 160~\micron.
The SAGE-SMC $5\sigma$ point source sensitivities, based on
completeness of the point source catalogs, are 0.045, 0.028, 0.12
and 0.10~mJy in the IRAC 3.6, 4.5, 5.8 and 8~\micron\ bands,
respectively (\S\ref{irac_completeness}), and 0.7, 25 and 200~mJy in
the MIPS 24, 70 and 160~\micron\ bands, respectively
(\S\ref{sec_mips_points}).  The 5$\sigma$ diffuse emission
sensitivities are $\sim$0.1, 0.3, 2.5, and 4 MJy/sr at IRAC 8.0, MIPS
24, MIPS 70, and MIPS 160~\micron, respectively.

\subsection{Observations}
\label{sec_obs}

\begin{figure*}[tbp]
\epsscale{1.1}
\plotone{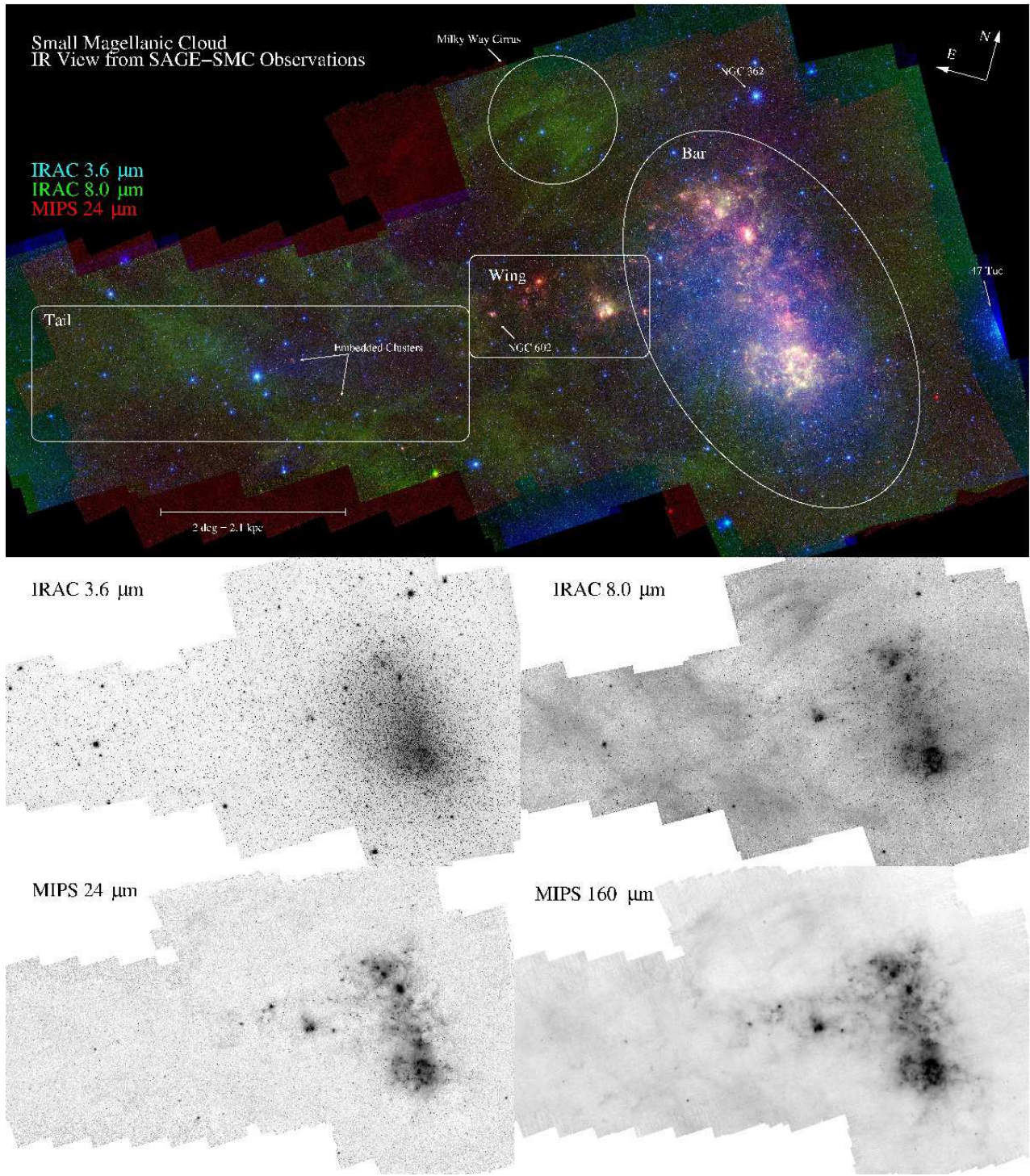}
\caption{The SAGE-SMC observations are illustrated in a 3-color image
and four single band images.  The 3-color image
gives the over all SAGE-SMC IR view of the SMC with annotations of the
Body, Wing, and Tail.  The locations of 47 Tuc and NGC 362 (two
Galactic globular clusters) are shown.  The green filamentary
structure seen throughout the image is due to Milky Way foreground
cirrus emission.  The single band images shown are the IRAC
3.6~\micron\ (stars), IRAC 8.0~\micron\ (aromatic emission), MIPS
24~\micron\ (hot dust/star formation), and MIPS 160~\micron\ (cold
dust).  All the images are displayed with a sinh$^{-1}$ stretch.
\label{fig_mosaics}}
\end{figure*}

The SAGE-SMC observations were taken with $1.1\arcdeg
\times 1.1\arcdeg$ degree tiles of IRAC High Dynamic Range (HDR) exposures, and 
MIPS fast scans with varying scan leg lengths.  To minimize artifacts
that limit 
sensitivity, we mapped at two epochs, separated by $\sim$3 (IRAC) and
$\sim$9 (MIPS) months, which provides a $\sim$90$\arcdeg$ roll
angle difference. This mapping strategy maximizes observing
efficiency while minimizing artifacts that compromise data
quality. The IRAC and MIPS artifacts fall in two 
classes: random effects (e.g.\ cosmic rays, bad pixels) and systematic
effects that are tied to pixel location and usually systematically
affect rows/columns. IRAC systematic effects include: saturation
latents, scattered light, MUX bleed, banding, and column
pulldown \citep{Hora04IRAC}. MIPS systematic effects include:
jailbars, pick-off mirror spots, latent images, and scattered light
at 24~\micron\ \citep{Engelbracht07MIPS24}, streaking due to saturation 
latents and time-dependent responsivity drifts at 70 \& 160~\micron\
\citep{Gordon07MIPS70},  
and insufficient 160~\micron\ coverage in fast scan mode. Clean
removal of random effects requires at least 4 overlapping
measurements. Systematic effects are optimally removed by combining
images taken with a $\sim$90$\arcdeg$ roll angle difference. This
type of strategy is recommended by the {\em Spitzer} Science Center (SSC) and the IRAC and MIPS 
instrument teams and has been proven to be very successful in the
SAGE-LMC data analysis \citep{Meixner06}.

To achieve the above goals, four 12s IRAC HDR exposures were taken in
pairs at two different epochs.  Each HDR exposure consists of both a
12s and 0.6s exposures and after accounting for deadtime the total
effective exposure time per pixel is 41.6 s.  A tile size of
$1.1\arcdeg \times 
1.1\arcdeg$ was executed in mapping mode using 120 pixel steps between
rows and 240 pixel steps between columns.  Steps were done instead of
dithers to minimize the time required to cover the whole area, and to
ensure no gaps occurred in the region being mapped.  Each IRAC
Astronomical Observing Request (AOR) consists of a 14$\times$28 map of
12s HDR frames. Two AORs overlap at each position, resulting in
coverage of at least four 12s HDR frames.

Each MIPS AOR consists of 8-16 fast scan legs that are
2$\arcdeg$--5$\arcdeg$ long with 1/2 array cross scan steps.  The SMC
was observed with an optimized grid of these
AORs.  Tight sequential constraints
relative to the roll 
angle rate of change were used so that neighboring long strips
had sufficient overlap.  We carefully designed our MIPS strategy
to ensure off-source measurements in every scan leg which allows
for accurate self-calibration of the instrumental effects. While MIPS
fast scan mode does not achieve full coverage at 160~\micron, the
SAGE-LMC observations have shown that the use of the two sets of
observations with one set rotated $\sim$90\arcdeg\ from the other
produces a mostly filled 160~\micron\ map \citep{Meixner06}.
The exposure times per pixel are 60s, 30s, and $\sim$9s at 24, 70, and
160~\micron, respectively.


\begin{deluxetable*}{lccccccc}
\tablewidth{0pt}
\tablecaption{{\em Spitzer} IRAC/MIPS Observations \label{tab_obs}}
\tablehead{\colhead{Origin} & \colhead{Instrument} & \colhead{Epoch} & \colhead{Dates} 
   & \colhead{Coverage} & \colhead{Depth} & \colhead{\# of AORs} &
  \colhead{\# images/band} \\ 
  & & & & \colhead{[$\sq\arcdeg$]} & \colhead{[sec]} & & }
\startdata
S$^3$MC  & IRAC & 0 & May 7-9, 2005   & 2.8 & 144 & 10 & $6.7 \times 10^3$ \\
SAGE-SMC & IRAC & 1 & Jun 12-19, 2008 & 30  & 20.8 & 29 & $1.1 \times 10^4$ \\
SAGE-SMC & IRAC & 2 & Sep 15-23, 2008 & 30  & 20.8 & 29 & $1.1 \times 10^4$ \\
S$^3$MC  & MIPS & 0 & Nov 6-8, 2004   & 3.7 & 80, 40, 16 & 7 & $1.2 \times 10^4$ \\
SAGE-SMC & MIPS & 1 & Sep 17-25, 2007 & 30 & 30, 15, 4.5 & 21 & $5.0 \times 10^4$ \\
SAGE-SMC & MIPS & 2 & Jun 25-28, 2008 & 30 & 30, 15, 4.5 & 24 & $4.7 \times 10^4$
\enddata
\end{deluxetable*}

The IRAC and MIPS observation dates and details are given in
Table~\ref{tab_obs}.  The details of the S$^3$MC observations are
given by \citet{Bolatto07} and included in Table~\ref{tab_obs} for
completeness.

\subsection{IRAC}

The SAGE-SMC IRAC data were processed with the Wisconsin pipeline that
also was used for the SAGE-LMC IRAC data \citep{Meixner06}.  The
pipeline starts with flux calibrated IRAC frames \citep{Reach05}
provided by the {\em Spitzer} Science Center (SSC).  These were processed
with SSC pipeline version S14.0 (S$^{3}$MC, epoch 0), S18.0.2 (epoch
1) and S18.1 (epoch 2). The Wisconsin pipeline removes or corrects for
image artifacts like cosmic rays, stray light, column pulldown, banding,
bad or missing pixels \citep{Hora04IRAC}; does point source extraction and bandmerging
across multiple observations and wavelengths; and mosaics the images.
During the band-merging process, the source lists are merged with a
combined All-Sky 2MASS \citep{Skrutskie06} and 6X2MASS point source
list \citep{Cutri032MASS} The pipeline produces two kinds of source lists:
a highly reliable Catalog and a more complete Archive.  Source lists
are made for each epoch of data from photometry on individual frames.
Source lists are also made from mosaic images for which all epochs
have been combined.  These combine single frame photometry at the
bright end with mosaic photometry at the faint.  Enhanced images (with
corrections for scattering light and image artifacts) and
residual images (point sources removed) of all epochs of data combined
are also produced.  Information more detailed than is given in this
paper about the processing and data products can be found in the
SAGE-SMC Data Description Delivery 3 document (hereafter
DDD3)\footnote{http://data.spitzer.caltech.edu/popular/sage-smc/}.

\subsubsection{Mosaic Images}

Both SAGE-SMC epochs of data and the S$^{3}$MC data were combined to
produce the mosaic images.  The images are
mosaicked using the
Montage\footnote{http://montage.ipac.caltech.edu/} package (v3.0), and
are projected to equatorial coordinates. 
The mosaic images conserve the surface brightness
from the original images; the units are MJy/sr. 
We matched instrumental background variations between the images using
Montage's level background correction algorithm.
The background-matching process introduced an artificial gradient in
the mosaic images which was removed by comparing the large-scale
background variations to the original images.

Residual images (images with point sources removed) were produced by
running DAOPHOT \citep{Stetson87} on the 12 second frame-time BCD
frames.  Note that we repeat the photometry calculations on the
residual, BCD images 
(referred to as ``tweaking''), which has been shown to substantially
improve the flux estimates in complex background regions.  The
residual mosaics are created from individual frame residual images.  Thus, if
a source is extracted in some but not all frames it will show up in
these images as a source (although its brightness will be reduced
because it is being averaged with images where the source was
extracted).  Sources may not be extracted for a variety of reasons,
mainly due to cosmic ray contamination, saturation/non-linearity
limits and along the frame edges.

\subsubsection{Point Source Lists}

There are two SAGE-SMC IRAC point source lists created named Catalog and
Archive.  The Catalog contains only the highly reliable sources while
the Archive is a more complete list both in number of sources and flux
measurements at each wavelength (less nulling of fluxes).  The main
differences between the Catalog and Archive are 1) fluxes brighter
than a threshold that marks a nonlinear regime are nulled (removed) in
the Catalog; 2) sources within 2\arcsec\ of another are culled
(removed) from the Catalog, whereas the Archive allows sources as
close as 0$\rlap.{''}$5 from another; 3) sources within the PSF
profile of a saturated source are culled from the Catalog but not the
Archive; and 4) the Catalog has higher signal-to-noise thresholds and
slightly more stringent acceptance criteria.  Table~\ref{tab_irac_cat}
lists the faint and bright limits and numbers of sources in the
various SAGE-SMC IRAC source lists.

\begin{deluxetable*}{lclrrrrr}
\tablewidth{0pt}
\tablecaption{IRAC Point Source Catalogs \label{tab_irac_cat}}
\tablehead{\colhead{Origin} & \colhead{Epoch} & \colhead{Type} 
   & \colhead{\#} & \multicolumn{4}{c}{Min/Max [mag]} \\
  & & & & \colhead{IRAC1} & \colhead{IRAC2} & \colhead{IRAC3} &
    \colhead{IRAC4} }
\startdata
S$^3$MC  & 0     & archive   &  272,716 & 6.0/17.7 & 5.5/17.1 & 3.0/14.7 & 3.0/14.1 \\
SAGE-SMC & 1     & archive   & 1,281,740 & & & & \\
SAGE-SMC & 2     & archive   & 1,177,258 & & & & \\ [0.1in]
S$^3$MC  & 0     & catalog   &  216,845 & 6.0/17.6 & 5.5/17.0 & 3.0/14.6 & 3.0/13.6 \\
SAGE-SMC & 1     & catalog   & 1,229,683 & & & & \\
SAGE-SMC & 2     & catalog   & 1,128,208 & & & & \\ [0.1in]
SAGE-SMC SMP & 0+1+2 & archive & 2,194,836 & 6.0/18.5 & 5.5/18.1 & 3.0/16.2 & 3.0/15.4 \\
SAGE-SMC SMP & 0+1+2 & catalog & 2,015,403 & 6.0/18.3 & 5.5/17.7 & 3.0/15.7 & 3.0/14.5
\enddata
\end{deluxetable*}

The single frame photometry source list fluxes were extracted from the
IRAC frames using a modified version of DAOPHOT to perform the PSF
fitting.  The point source lists
are merged at two stages using a modified version of the SSC
bandmerger:
first a merge of all detections at a single wavelength and different
observation times; and then a merge of all wavelengths at a given
position on the sky.  Stringent selection criteria are then applied to
each source to ensure that the delivered lists contain only legitimate
astronomical sources with high-quality fluxes.  The point source
extraction, bandmerging steps, and catalog source selection criteria
are described in much more detail in the DDD3 document.

The Single Frame + Mosaic Photometry (SMP) Catalog and Archive are a
combination of mosaic photometry and the single frame photometry Epoch
0+1+2 source lists.  The single-frame photometry produces more
accurate fluxes at the bright end due to its well-defined
point-spread-function.  Epoch 0+1+2 source lists were derived from
doing photometry on individual IRAC frames, then doing an
error-weighted average of those results for each band.  The mosaic
photometry is done on the 12 second frametime mosaiced images (mosaics
of combined Epoch 0+1+2) which have been cleaned of most of the
instrument artifacts, including cosmic rays (CRs) (which are abundant
in the single frames).  The mosaic photometry and catalog generation
are described in a SAGE-LMC
document\footnote{http://data.spitzer.caltech.edu/popular/sage/20090922\_enhanced/\\documents/SAGEDataProductsDescription\_Sep09.pdf}
and will also be described in a forthcoming SAGE-SMC data delivery document.

\subsubsection{Precision and Accuracy of the Photometry}

\begin{figure*}[tbp]
\epsscale{1.0}
\plotone{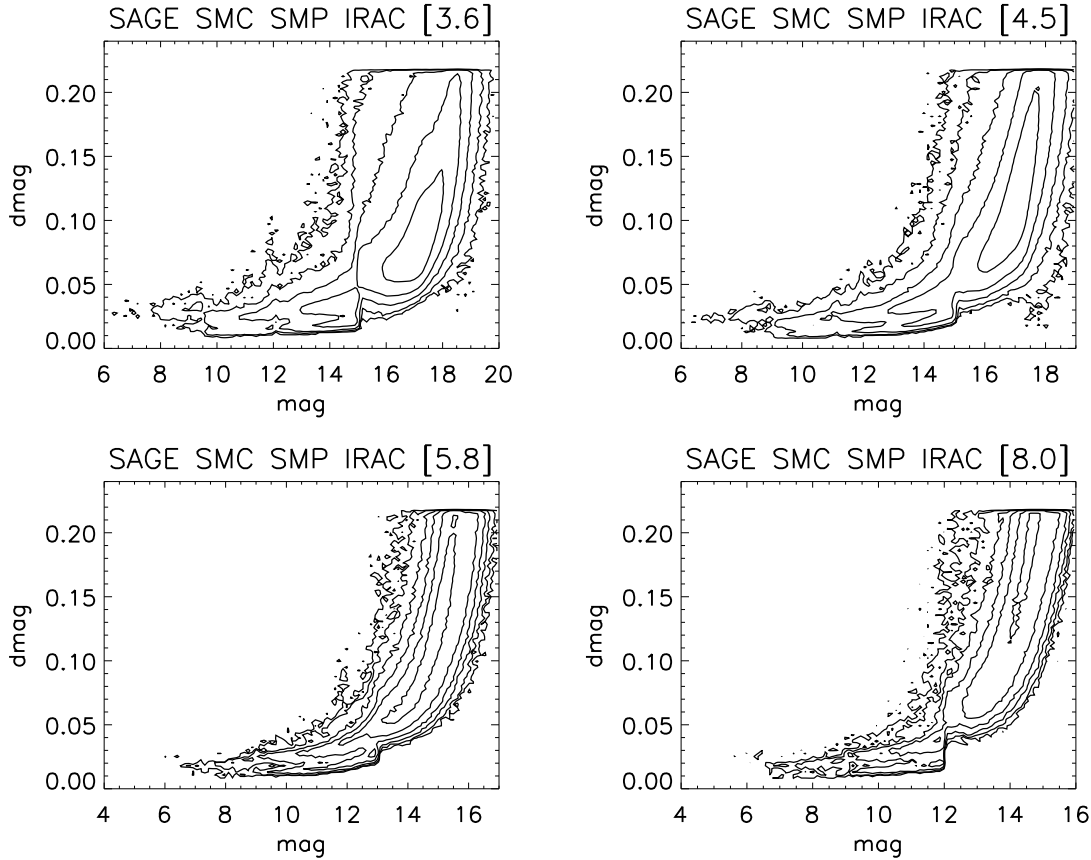}
\caption{Magnitude uncertainty versus magnitude for each IRAC band
include in the SAGE-SMC SMP Archive.  Contours show the
density of the sources.  The lack of data above dmag of 0.22 is caused
by the criterion that archive data have signal to noise ratios of 5 or
better. \label{fig_irac_cat_unc}}
\end{figure*}

Figure~\ref{fig_irac_cat_unc} shows the photometric uncertainty for
the SAGE-SMC SMP Archive. There is a jump in uncertainties at the
brighter magnitudes, e.g. 9.5 at 3.6 $\mu$m, which shows the boundary
between the 0.6 and 12 second photometry (with the shorter exposure
having larger errors).  In addition, there is a discontinuity in the
uncertainties at the magnitude where the photometry transitions
between single-frame to mosaic.  We choose to use the mosaic
photometry results over the single-frame photometry (when applicable)
to avoid the known problem of Malmquist bias in our single-frame
photometry as our single-frame photometry approaches it's faint limit
whereby the single-frame photometry is known to become progressively
too bright.  We believe these mosaic photometry values are more
reliable (although with somewhat larger uncertainties) than the
single-frame photometry results

To ensure that our photometric calibration is uniform across the large
area observed by SAGE-SMC, and between different AORs, epochs, and
wavelengths, we compared our photometry to a network of absolute
stellar calibrators custom-built for SAGE-SMC. These are 53 A0-A5V or
K0-M2III stars selected from SIMBAD; their surface density within the
SAGE SMC area is approximately 3 stars/$\sq\arcdeg$, uniformly
distributed across the SMC.  The
predicted fluxes of the calibrators are tied to the {\em MSX} absolute
calibration which has an accuracy of 
$\pm$1.1\% \citep{Price04}. The method employed to produce the
SAGE-SMC network of calibrators is identical to that used to create
the suite of standards at the North Ecliptic Pole \citep{Cohen03IRAC}
from which the IRAC primary calibrators were selected \citep{Reach05}.

The agreement between the SAGE-SMC magnitudes and those predicted from
the calibration stars is excellent (plots can be found in the DDD3
document, \S3.5.2), with differences between the two much smaller than
the one-sigma errors of our photometry. The ensemble averaged
differences and standard deviations in the four IRAC bands between
SAGE-SMC and the predicted magnitudes are 0.010$\pm$0.062,
0.024$\pm$0.060, --0.002$\pm$0.060, --0.018$\pm$0.052, for bands [3.6],
[4.5], [5.8], and [8.0] respectively, for Epoch 1 and are similar to
those for Epoch 2.  These values are
consistent with other surveys processed with the Wisconsin IRAC
pipeline (e.g. GLIMPSE and SAGE-LMC). There is no statistical offset
between the predicted magnitude of the calibrators and the extracted
values.

The S$^3$MC observations (epoch 0) were re-processed as part of the
SAGE-SMC project.  The SAGE-SMC epoch 0 IRAC photometry of point
sources showed systematic offsets when compared to the S$^{3}$MC
catalog \citep{Bolatto07} in the [5.8] and [8.0] bands ($\sim$0.2~mag;
see \S3.5.4 of the DDD3).  The S$^3$MC region contains 6 of the 53 primary flux
calibrators described in the previous section and the SAGE-SMC epoch 0
archive magnitudes of the 6 calibrators shows agreement to within the
formal photometric errors.  This comparison suggests that the
Wisconsin IRAC pipeline processing produces no systematic offsets from
magnitudes predicted by the flux calibrators (at least down to 11th
magnitude, the faintest calibrators).  The origin of the differences
is not specifically known, but it is suspected to be somewhere in
the mosaicking and photometry stage of the S$^3$MC processing.

The consistency of the SAGE-SMC IRAC photometry was checked by
comparing photometry of sources found in the SAGE-SMC epoch~0 and
epoch~1 catalogs. There is no offset and the degree of scatter,
especially at the brighter magnitudes, is likely due to true variables
in these two data sets (see \S3.5.4 of the DDD3).  Since the SAGE-SMC
data agree well with the SAGE-SMC flux calibrators, we conclude that
the Wisconsin processing of the S$^{3}$MC data is well calibrated.

\subsubsection{Completeness}
\label{irac_completeness}

\begin{figure*}[tbp]
\plotone{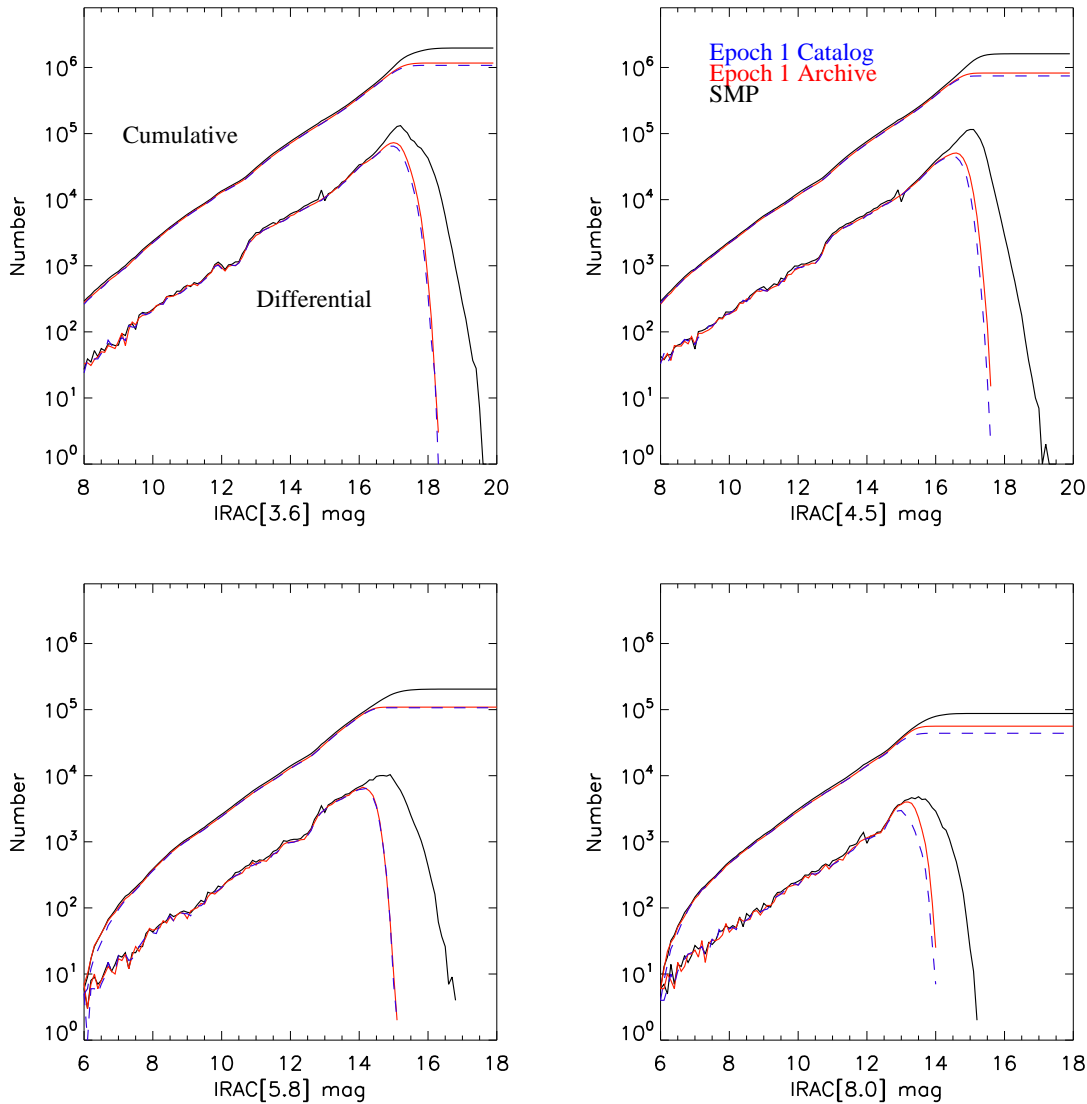}
\caption{Source counts as a function of magnitude for the SAGE-SMC
epoch 1  Catalog (dashed blue line), 
the SAGE-SMC epoch 1 Archive (red line),  and
the SAGE-SMC single frame+mosaic photometry Archive (solid black line).
The lower three lines
in each panel show the differential number counts while the upper three
lines in each panel show the cumulative number counts. \label{fig_irac_counts}}
\end{figure*}

In Figure~\ref{fig_irac_counts} we compare the number counts per
magnitude bin of sources in the epoch 1 Archive, the epoch 1 Catalog,
and the single frame + mosaic photometry Archive of the combined
datasets.  The epoch 1 lists are complete down to 16.0, 15.5,
13.0 and 12.0 mag in IRAC bands [3.6], [4.5], [5.8], and [8.0]
respectively with the big drop-offs at 17.0, 16.5, 14.0 and 13.5 mag,
respectively.  Completeness is also a function of background level
which is more variable for IRAC [5.8] and [8.0].  We note that the
single-frame Archive and Catalog source lists were designed for
reliability over completeness.

The single frame + mosaic photometry source lists extend the Archive
and Catalog about 1.5 magnitude deeper than the single-frame
photometry single-epoch source lists.  These lists are mostly complete
down to 16.5, 16.5, 14.5 and 13.5 in IRAC bands [3.6], [4.5], [5.8],
and [8.0] respectively; with big drop-offs at 17.0, 17.0, 15.0 and
14.5, respectively.  These sourcelists are more complete due to cosmic
ray removal from the mosaic images before source extraction, and the
lower noise from the mosaic images.

\subsection{MIPS}
\label{sec_mips}

The MIPS data were processed with the MIPS Data Reduction Tool \citep[DAT,][]{Gordon05DAT} and
customized post-processing scripts.  This processing was similar to
that done for the SAGE-LMC data \citep{Meixner06}, with some
updates that have also been applied to the final SAGE-LMC data
deliveries.  At 24~\micron, the 
custom post-processing steps (for each AOR) included masking of bright
source latents, a 2nd scan mirror independent flat field, background
subtraction (mainly removing the variable zodiacal light), correction of
piecewise jailbars produced by bright 
saturating sources, matching the levels between adjacent images (when
offsets were found to be above the noise), and correcting for scan
mirror position dependent scattered light.  The custom post-processing
steps at 70~\micron\ are the removal of the instrumental baseline
variations and correction for the flux dependent non-linearities.  The
instrumental baseline variations are removed by subtracting a low
order polynomial fit as a function of observation time to the data taken 
outside of the SMC.  This step is only possible as the observations
were designed such that each scan leg included measurements beyond the
SMC.  The flux dependent non-linearities were removed using a
correction determined from the 70~\micron\ calibration stars
\citep{Gordon07MIPS70}, specifically a comparison of the observed
surface brightnesses to those predicted from the fitted PSF.
At 160~\micron, the custom post-processing steps were to remove
cosmic rays using a spatial filter and interpolate over the few, single
pixel sized holes in the final combined epoch coverage.  The SAGE-SMC
mosaics created from the
combined S$^3$MC (corrected, see \S\ref{sec_s3mc_cor}) and SAGE-SMC
observations are 
shown in Fig.~\ref{fig_mosaics}. 

\subsubsection{Point Source Catalogs}
\label{sec_mips_points}

Point source catalogs were created for all three MIPS bands using the
PSF-fitting program StarFinder \citep{Diolaiti00} using model PSFs
validated as part of the MIPS calibration effort
\citep{Engelbracht07MIPS24, Gordon07MIPS70, Stansberry07MIPS160}.
The StarFinder photometry package is optimized for images with well
sampled PSFs such as that of 3 MIPS bands.  A different photometry
package optimized for undersampled images was used for IRAC.  The
construction of the catalogs is described below and summarized in
Table~\ref{tab_mips_cat}.  The flux and magnitude limits are for
regions with the least confusion either due to source crowding or
complex backgrounds.

\begin{deluxetable*}{lcclrrrrr}
\tablewidth{0pt}
\tablecaption{MIPS Point Source Catalogs \label{tab_mips_cat}}
\tablehead{\colhead{Origin} & \colhead{Band} & \colhead{Epoch} & \colhead{Type} 
   & \colhead{\#} & \colhead{Min} & \colhead{Max} & \colhead{Min} & \colhead{Max} \\ 
  & & & & & \multicolumn{2}{c}{Flux [mJy]} & \multicolumn{2}{c}{[mag]} }
\startdata
S$^3$MC  & MIPS24  & 0     & full     & 18,067 & 0.09 & $6.9\times10^{3}$ & 12.2 & 0.04 \\
         &         &       & high-rel &  3,657 & 0.77 & $5.7\times10^{3}$ & 9.9 & 0.25 \\ 
SAGE-SMC & MIPS24  & 1     & full     & 67,068 & 0.13 & $6.9\times10^{3}$ & 11.9 & 0.05 \\
         &         &       & high-rel & 12,974 & 0.72 & $6.2\times10^{3}$ & 10.0 & 0.16 \\  
SAGE-SMC & MIPS24  & 2     & full     & 63,102 & 0.12 & $6.6\times10^{3}$ & 12.0 & 0.09 \\
         &         &       & high-rel & 13,071 & 0.72 & $6.6\times10^{3}$ & 10.0 & 0.09 \\
SAGE-SMC & MIPS70  & 0+1+2 & full     &  6,123 &  8.4 & $4.0\times10^{4}$ & 4.9 & --4.3 \\
+S$^3$MC &         &       & high-rel &    939 &   21 & $2.0\times10^{4}$ & 3.9 & --3.5 \\ 
SAGE-SMC & MIPS160 & 0+1+2 & full     &    953 &   45 & $2.5\times10^{4}$ & 1.4 & --5.5 \\
+S$^3$MC &         &       & high-rel &    132 &  150 & $2.0\times10^{4}$ & 0.10 & --5.2 
\enddata
\end{deluxetable*}

\begin{figure*}[tbp]
\epsscale{1.1}
\plotone{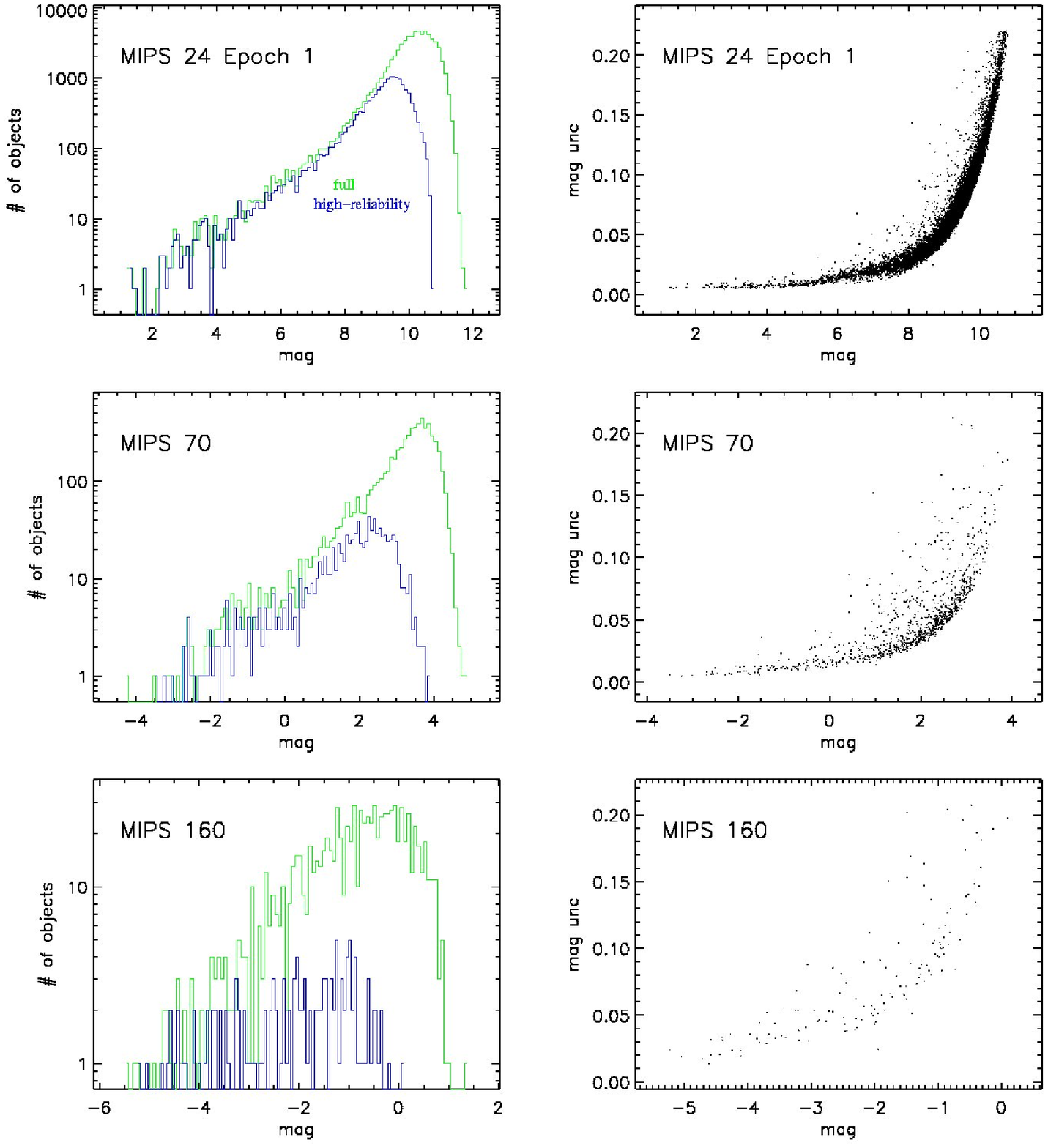}
\caption{The magnitude histograms for the ``full'' and ``high-reliability''
MIPS 24~\micron\ epoch 1, 70~\micron, and 160~\micron\ catalogs are shown on the left.  On the
right, the magnitude versus uncertainty plot for same
``high-reliability'' catalogs are plotted.  The MIPS 24~\micron\ epoch 0 and epoch 2
plots are very similar to the MIPS 24~\micron\ epoch 1 plots.
\label{fig_all_cat}}
\end{figure*}

At 24~\micron, the catalogs were created from the individual AOR
mosaics and the individual AOR point source catalogs were merged to
create the ``full'' catalog.  Only sources with StarFinder correlation
thresholds above 0.80 and $\ge 2\sigma$ were included in the ``full''
catalog.  The StarFinder correlation parameter quantifies how close
each source is to the input PSF where a value of 1 is a perfect match.
The $2\sigma$ cut was done using the uncertainty image 
created during the mosaicking of the individual images.  The goal of
the ``full'' catalog is to produce as complete a 
catalog as possible to the faintest levels.  The ``full'' catalog
contains sources known to be false.  A ``high-reliability'' catalog is
created from the ``full'' catalog by removing all sources with a
StarFinder correlation $<$0.89, sources with signal-to-noise $<
5\sigma$, and sources with a StarFinder
correlation $<$0.91 in regions where the background is highly
structured (background standard deviation greater than 0.25).  These cuts were
determined by extensive visual testing of the AOR catalogs against the
AOR mosaics.  This removes most false sources, but at the expense of
removing real sources as well.  The cut at a StarFinder
correlation of 0.89 removed most of the sources with the other two
cuts removing many fewer sources.  Catalogs were produced for all
the epochs at MIPS 24~\micron.  The flux histograms for both
epoch 1 catalogs and the flux versus signal-to-noise plot are shown in
Fig.~\ref{fig_all_cat}.

At 70 and 160~\micron, the point source catalogs are created from the
combined mosaics.  This is required due to the need to suppress
residual instrumental artifacts present in the single epoch mosaics
(mainly at 70~\micron) and have sufficient coverage (mainly at
160~\micron).  The 70 and 160~\micron\ ``full'' and ''high-reliability''
catalogs were generated using StarFinder with the same settings as for
24~\micron.  The flux histograms for these
catalogs and the flux versus signal-to-noise plots are shown in
Fig.~\ref{fig_all_cat}.

We compared the 24~\micron\ and 70~\micron\ fluxes we measured using
the S$^3$MC observations (i.e., epoch 0) versus those presented by
\citet{Bolatto07} measured from the same data.  Out of the 16130
(1762) sources from \citet{Bolatto07} 
that have 24 (70~\micron) fluxes, we find matches in
our epoch 0 full catalog for 9878 (265) sources within $4\arcsec$
($20\arcsec$).  The ratio of fluxes between the S$^3$MC/SAGE-SMC
measurements are $1.015 \pm 0.0017$ and $0.913 \pm 0.025$ for 24 and
70~\micron, respectively.  The difference at 24~\micron\ is due to
updated processing and calibration factor after after the S$^3$MC
catalog was created \citep{Engelbracht07MIPS24}.   The difference at
70~\micron\ is due to a 
combination of uncorrected 70~\micron\ nonlinearities in the S$^3$MC
catalog and the 11\% change in the 70~\micron\ calibration factor
after the S$^3$MC catalog was created \citep{Gordon07MIPS70}.

\subsection{Legacy Data Products}

One component of a {\em Spitzer} Legacy project is to provide a legacy to
the entire community.  The first step is the SAGE-SMC raw and SSC
pipeline reduced data have no proprietary period.  Thus, these data
are available to the community and SAGE-SMC team at the same time.
The next step is to provide higher level data products to the
community.  For the SAGE-SMC project, these higher level data products
are the point source catalogs and mosaicked images of the entire SMC.
These higher level products are available through the {\em Spitzer}
Science Center and IRSA Archive.  We expect these products to be
useful for a number of different science topics as well as followup
observations (e.g., from the ground, the {\em Herschel Space Observatory},
the {\em James Webb Space Telescope}, etc.).  The SAGE-SMC processed version
of the S$^3$MC data (catalogs and mosaics) are included in the
SAGE-SMC deliveries.

\section{Results}

\subsection{Nomenclature}

The multiwavelength appearance of the SMC (Fig.~\ref{fig_smc_mwave})
clearly shows this galaxy is made of three main components: the Bar,
Wing, and high-density portion of the Magellanic Bridge that we
refer to as the Tail.   The nomenclature for parts of the SMC has
varied somewhat in the literature.  In order to be clear, we give the 
definitions we use and how they relate to past definitions.  

The regions are defined pictorially in Fig.~\ref{fig_mosaics}.  The SMC
Bar is the region that contains the majority of the older stars and star
formation in the SMC.  The SMC Wing extends eastward from the Bar and contains
clear strong star formation ($1\fh 08^m \la$ RA $\la 1\fh 31^m$).  The SMC Tail
extends eastward from the Wing ($1\fh 31^m \la$ RA $\la 2\fh 30^m$) and
corresponds to the high density portion of the Magellanic 
Bridge \citep{Muller03}.  Previous definitions of the SMC Wing have
also included all of the Tail. Our specific definitions of the Wing
and Tail are motivated as the two regions display significantly
different characteristics.  The SMC
Wing includes large well known \ion{H}{2} regions (e.g., NGC 602),
high density \ion{H}{1} (Fig.~\ref{fig_smc_mwave}), and old stars.
The SMC Tail has some young stars \citep{Harris07} and embedded star
formation regions \citep{Gordon09SMC}, but no old stars and fairly low
\ion{H}{1} column densities.  In fact, the SMC Tail has the
characteristics of a tidal tail, which has been recently pulled out of
the main SMC body.  The tidal nature of the SMC Tail is supported by
the fact that it has the same metallicity \citep{Rolleston03, Lee05}
and gas-to-dust ratio \citep{Gordon09SMC} as the SMC Body.

\subsection{Global Infrared SED}

\begin{deluxetable*}{lccccc}
\tablewidth{0pt}
\tablecaption{SMC Global Fluxes \label{tab_global_sed}}
\tablehead{ & \colhead{Wavelength} & \colhead{Bandwidth} & \colhead{Flux} & & \\
 \colhead{Band} & \colhead{[\micron]} & \colhead{[\micron]} & \colhead{[kJy]} & 
 \colhead{Origin} & \colhead{Reference} }
\startdata
IRAC1   & 3.550 & 0.681 & $0.30 \pm 0.02$  & {\em Spitzer}/IRAC & 1 \\
IRAC2   & 4.493 & 0.872 & $0.22 \pm 0.01$  & {\em Spitzer}/IRAC & 1 \\
IRAC3   & 5.731 & 1.250 & $0.22 \pm 0.01$  & {\em Spitzer}/IRAC & 1 \\
IRAC4   & 7.872 & 2.526 & $0.20 \pm 0.01$  & {\em Spitzer}/IRAC & 1 \\
{\em IRAS}12  & 12    & 7.0   & $0.13 \pm 0.04$   & {\em IRAS} & 2 \\
COBE12  & 12    & 6.48  & $0.08 \pm 0.03$   & COBE/DIRBE & 3 \\
MIPS24  & 23.7  & 4.7   & $0.35 \pm 0.01$  & {\em Spitzer}/MIPS & 1 \\
{\em IRAS}25  & 25    & 11.2  & $0.37 \pm 0.05$  & {\em IRAS} & 2 \\
COBE25  & 25    & 8.60  & $0.46 \pm 0.18$  & COBE/DIRBE & 3 \\
{\em IRAS}60  & 60    & 32.5  & $7.45 \pm 0.25$  & {\em IRAS} & 2 \\
COBE60  & 60    & 27.84 & $8.45 \pm 0.37$  & COBE/DIRBE & 3 \\
MIPS70  & 71.0  & 19.0  & $11.2 \pm 0.50$ & {\em Spitzer}/MIPS & 1 \\
{\em IRAS}100 & 100   & 31.5  & $12.7 \pm 0.6$ & {\em IRAS} & 2 \\
COBE100 & 100   & 32.47 & $15.8 \pm 2.72$ & COBE/DIRBE & 3 \\
COBE140 & 140   & 39.53 & $14.0 \pm 5.6$   & COBE/DIRBE & 3 \\
MIPS160 & 156   & 35.0  & $21.7 \pm 2.6$ & {\em Spitzer}/MIPS & 1 \\
COBE240 & 240   & 95.04 & $9.6 \pm 4.4$    & COBE/DIRBE & 3 \\
TOPHAT  & 476   & 48    & $3.2 \pm 0.81$  & TopHat & 4 \\
TOPHAT  & 652   & 65    & $1.62 \pm 0.29$  & TopHat & 4 \\
TOPHAT  & 750   & 75    & $0.95 \pm 0.19$  & TopHat & 4 \\
TOPHAT  & 1224  & 306   & $0.32 \pm 0.08$  & TopHat & 4
\enddata
\tablerefs{(1) this work; (2) \citet{Wilke04};
 (3) color-corrected \citet{Stanimirovic00}; (4) \citet{Aguirre03}}
\end{deluxetable*}

The total fluxes in the IRAC and MIPS bands were measured using
circular apertures visually centered on the SMC with a radius of
$2.25\arcdeg$ and a background annulus from $2.3\arcdeg$ to
$2.5\arcdeg$.  The IRAC 
absolute calibration uncertainties are taken as 5\% which is larger
than the point source uncertainty of $\sim$2\% \citep{Reach05} as the
IRAC scattered light behavior makes the extended source calibration
less accurate.  The MIPS absolute uncertainties are 4\%
\citep{Engelbracht07MIPS24}, 5\% \citep{Gordon07MIPS70}, and 12\%
\citep{Stansberry07MIPS160} for the MIPS 24, 70, and 160~\micron,
respectively.  The new IRAC and MIPS fluxes as well as 
previous IR measurements of the total SMC emission are tabulated in
Table~\ref{tab_global_sed}.  An even more extensive table of existing
global SMC measurements is given by \citet{Israel10}.  Our IRAC and
MIPS measurements are consistent with previous measurements including
the MIPS global measurements by \citet{Leroy07SMC}.

\begin{figure}[tbp]
\epsscale{1.2}
\plotone{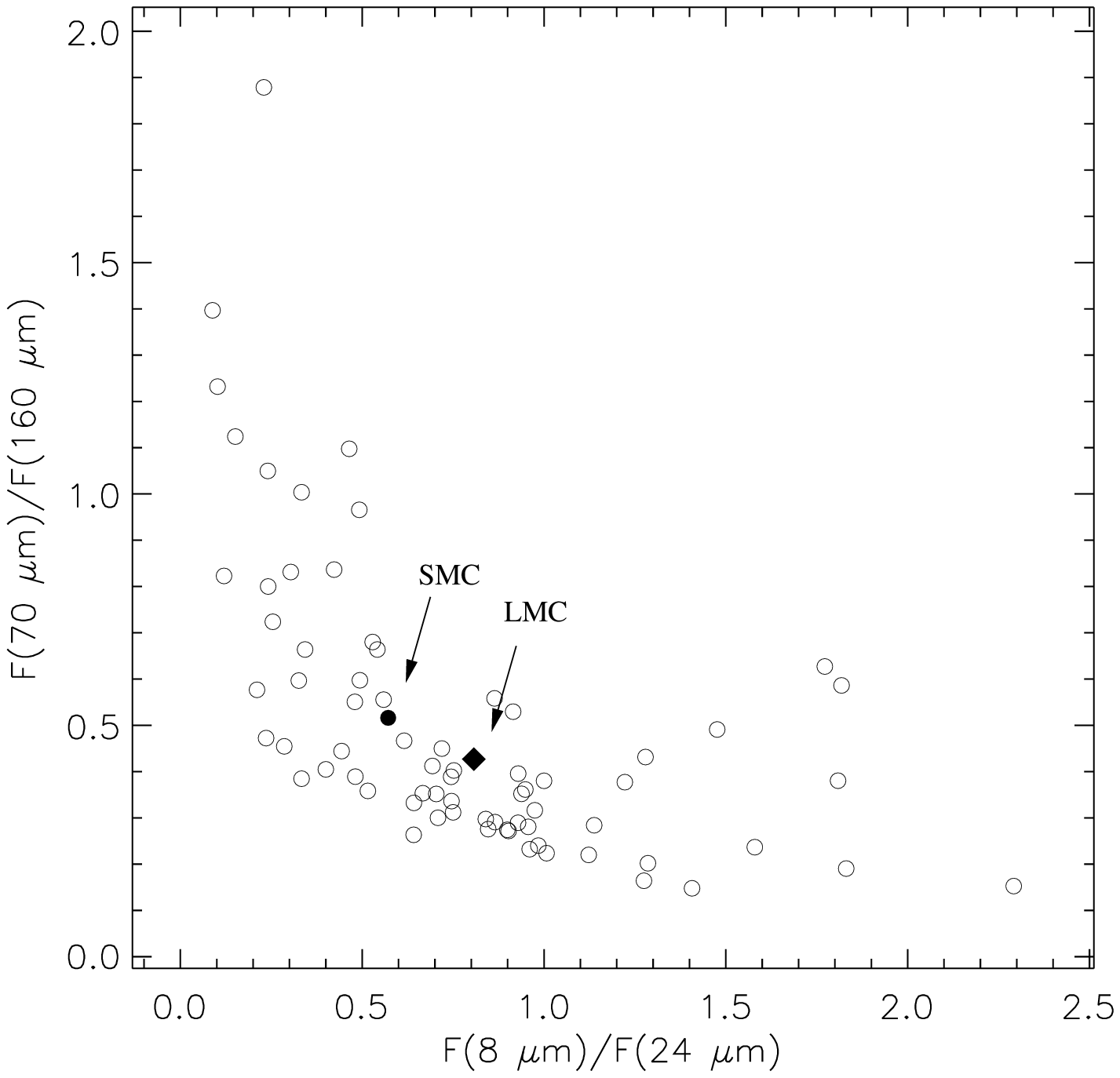}
\caption{The 8/24 and 70/160 infrared colors of the SMC and LMC are
plotted along with the same colors from the SINGS galaxy sample
\citep{Dale07}.
\label{fig_smc_sings_ann}}
\end{figure}

The infrared colors of the SMC are compared to those seen in the
SINGS sample of galaxies \citep{Dale07} in
Fig.~\ref{fig_smc_sings_ann}.  The location of the LMC is also
shown in this plot \citep{Bernard08}.  The location of the SMC in this
color--color plot indicates that it has somewhat warmer dust (higher
70/160 ratio) and lower aromatic/very hot dust emission (8/24
ratio) than the average SINGS galaxy and the LMC.  Overall the SMC
(and LMC) lie solidly in the middle of the SINGS galaxies IR colors.
This shows that the Magellanic Clouds are reasonable representations
of the average properties of more distant galaxies in the IR.  Thus,
they provide excellent places to study at high {\it physical}
resolution the details of the interestellar medium, star formation,
and stellar populations and how they relate to the global properties a
galaxy.  This is well illustrated by the recent \citet{Lawton10} work
that studied the large scale star formation properties of regions in
both Clouds.  They determined that the optimal monochromatic IR
obscured star formation rate indicator for physical scales larger than
10~pc is the MIPS 70~\micron\ band and that the average size of star
forming regions is $\sim$300~pc.  While using the Magellanic Clouds to
understand more distant galaxies is not new, our finding firmly
establishes this can be done in the IR.

\subsection{Interstellar Medium}

The SMC presents a distinct mix of ISM components different from that
found in the Milky Way (MW) and LMC.  For example, the molecular phase in the MW
dominates the inner disk and atomic gas dominates elsewhere, while the
diffuse ISM only has $\sim$15\% of the gas mass. In contrast, in the
SMC, the ionized ISM dominates, then the atomic gas and, finally, the
molecular ISM which is relatively confined and lower in mass
\citep{Leroy07SMC}.  The differences seen in the SMC are likely related
to its low metallicity at around 1/5 Z$_{\sun}$
\citep[Bar/Wing,][]{Russell92}.  Observations with {\em ISO}
\citep{Madden06} and {\em Spitzer} \citep{Engelbracht05SB, Engelbracht08SB,
Gordon08} have revealed that the ISM in low-metallicity environments
has weak/absent aromatic emission.  The aromatic emission is usually
attributed to PAH molecules/grains and the absence of PAHs has a profound
influence on the gas heating and the existence of cold/warm phases in
the ISM \citep{Wolfire95}. In particular, variations in the small
grain properties, as traced by the aromatic emission, are of fundamental
importance to the ISM thermodynamics since these grains are efficient
in heating the gas through the photoelectric effect \citep{Bakes94, vanLoon10}.

Previous observations in the SMC have shown large variations in dust
properties: dust in the Bar has very weak aromatic features and has UV
extinction with a steep UV rise and no 2175~\AA\ bump, while the dust
in the Wing shows MW-like UV extinction and aromatic features
\citep{Gordon98, Li02, Gordon03, Bolatto07, Sandstrom10}.  Additionally,
the gas-to-dust ratio has been seen to vary spatially across the SMC
by a factor of a few \citep{Gordon03, Bot04, Leroy07SMC}.  Understanding how
the dust varies in detail across the SMC is crucial to increasing our
understanding of dust in general and for galaxy evolution and star
formation as SMC dust is often used as a template for dust in
starburst and high-redshift galaxies \citep{Pei99}.  

The far-infrared 70 \& 160~\micron\ observations trace the dust
column density and, combined with the interferometric HI
\citep{Muller03, Stanimirovic04} and CO \citep{Mizuno01, Mizuno06}
observations, measures the gas-to-dust variations across the whole
SMC.  Preliminary results from SAGE-SMC observations in the SMC Tail find a
gas-to-dust ratio that is consistent with that found in the rest of
the SMC \citep{Gordon09SMC}.  This confirms that the SMC Tail is a tidal
feature consisting of SMC ISM and the discovery of YSOs confirms star
formation in situ (Chen et al., in prep.). 

\begin{figure}[tbp]
\epsscale{1.2}
\plotone{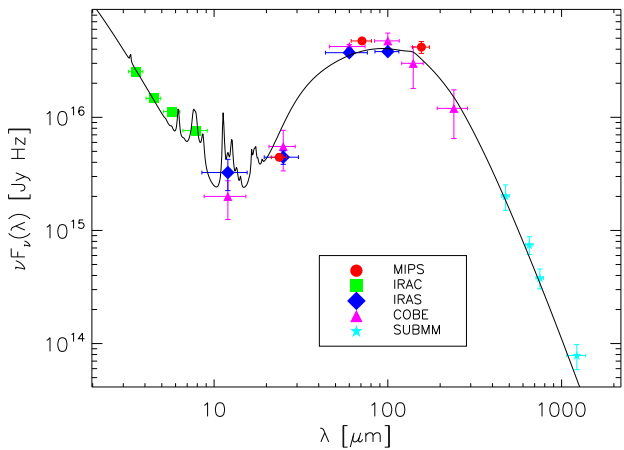}
\caption{The integrated global SED of the SMC is shown.  The best fit
  \citet{Draine07} model is given as the solid line.
\label{fig_smc_sed}}
\end{figure}

The integrated infrared SED of the SMC is plotted
Fig.~\ref{fig_smc_sed}.  The global SED was fit
with a model equivalent to that used by \citet{Draine07}.  The best
fit model was for Milky Way dust with
$R(V)=3.1$, a PAH mass fraction of 1.22\%, a f(PDR) = 0.25, and U values
(radiation field density normalized to the MW value) from 0.1--$10^3$. 
This model puts the SMC in the set of galaxies with 
low PAH abundances \citep[see Fig.\ 20 of][]{Draine07} which is not
surprising given the SMC's low metallicity [12+log(O/H) $\sim 8.0$
and, thus, high radiation field hardness \citep{Gordon08}.  The dust
mass implied by this fit is very high, $M(dust) = 2.1\times 10^6$~M$_\sun$, due
to the inclusion of the submm data points in the fit.  The
implications of the submm data in such fits of the SMC and
alternatives to large dust masses have been discussed at length by
\citet{Bot10} and \citet{PlankMCs11}.

The SMC dust properties are explored spatially in detail in the
companion SAGE-SMC ISM paper (Bot et al. 2011, in prep.). 

\subsection{Evolved Stars}

\begin{figure*}[tbp]
\epsscale{1.1}
\plotone{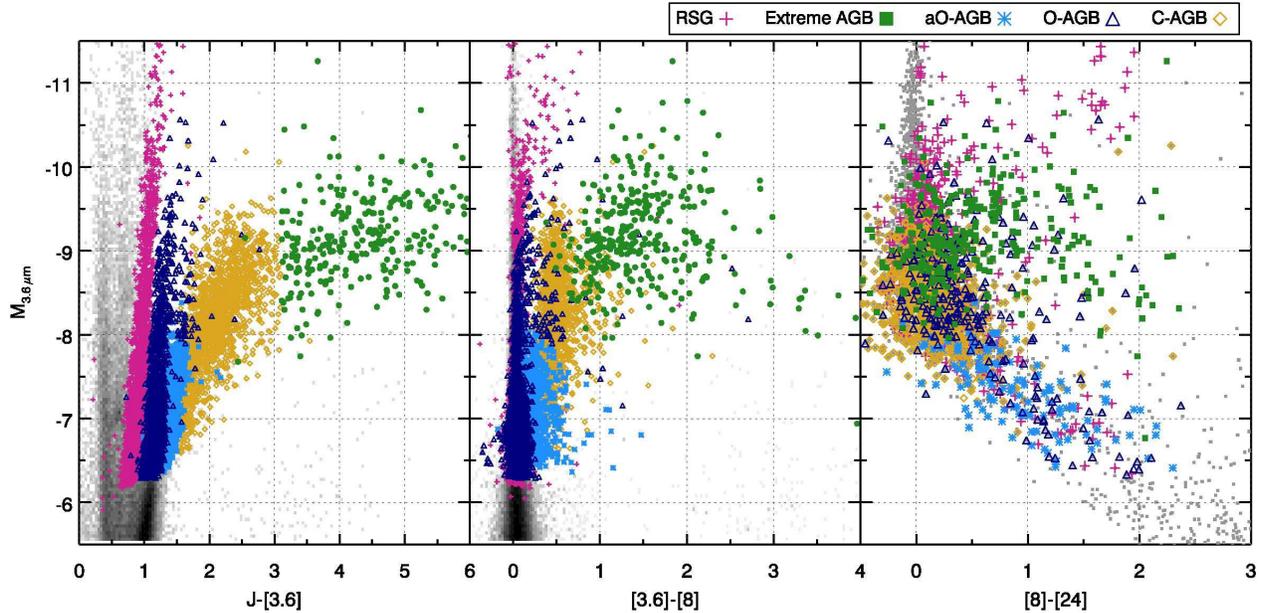}
\caption{M$_{\mathrm{3.6~\micron}}$ vs.\ $J-[3.6]$ ({\it left}), vs.\
$[3.6]-[8.0]$ ({\it middle}), and vs.\ $[8.0]-[24.0]$ ({\it right})
color--magnitude diagrams for the SMC are shown created using the IRAC SMP
Archive and and MIPS 24~\micron\ epoch 1 full list.
M$_{\mathrm{3.6~\micron}}$ is the absolute magnitude at 3.6~\micron\
and the [3.6], [4.5], [8], \& [24] labels give the Vega magnitude at
those wavelengths.  All sources are displayed as grayscale Hess
diagrams in the left and middle panels or gray points in the right
panel.  Sources were classified photometrically by their $J-K_{\rm S}$
colors, with the exception of anomalous O-rich AGB stars
(aO-AGB), which were classified by their $J-[8]$ color. RSGs are shown
as magenta crosses, O-rich AGB (O-AGB) stars as dark blue triangles,
aO-AGB stars as light blue asterisks, C-rich AGB stars (C-AGB) as
light orange diamonds, and extreme AGB stars (x-AGB) as green
dots. Few of the O-rich sources are detected at 24~\micron\, leaving
the C-AGB and x-AGB stars to dominate at 24~\micron\ point-source
flux.
\label{fig_smc_cmd_es}}
\end{figure*}

Heavy mass loss during the AGB and RSG phases, and for more massive
stars during the Wolf-Rayet and LBV phases, leads to the formation of 
circumstellar envelopes that are observable via their dust emission at
8~\micron\ and longer.  The SMC has significant metallicity and age
variations in the evolved star population over its full field of view
\citep{Harris04, Cioni06}.  The SAGE-SMC survey is sensitive to all
dust mass-losing evolved stars (mass-loss rates $> 10^{-8}$ M$_{\sun}$
yr$^{-1}$) across the entire SMC.

Stellar mass loss can drive the late stages of stellar evolution, yet
the mechanism for mass loss remains poorly understood.  The SAGE-SMC
survey quantifies the mass-loss rates by detecting excess emission at
8 and 24~\micron.  SAGE-SMC has detected numerous lower luminosity AGB
stars, the IR bright stars at the tip of the AGB (both C-rich and
O-rich), the "extreme" or obscured AGB stars with prodigious mass-loss
\citep{Blum06}, and the rare RSGs and other evolved stars
\citep{Bonanos10}.  Present 
estimates disagree on the relative contributions from these different
stellar classes to the injected mass budget of a galaxy
\citep{Tielens01}.  Modeling of the evolved star SEDs measured in the
SAGE-LMC survey has been successful both using empirical methods
\citep{Srinivasan09} and radiative transfer calculations
\citep{Srinivasan10, Sargent11}.  The results from SAGE-LMC work indicate the current
mass loss in the LMC is comprised of 24\% from each of the optically
visible C-rich and O-rich groups (C-AGB, O-AGB), 14\% from bright RSG
stars, and 37\% from extreme AGB (x-AGB) stars with a total dusty
mass-loss return of about $0.7 \times 10^{-2}$ M$_{\sun}$ yr$^{-1}$.
In the lower metallicity of the SMC, we find a higher fraction of
C-rich stars (C-AGB and x-AGB stars: 35.8\% in the SMC and 29.8\% in
the LMC) and hence a different distribution of mass-loss contributions
over the classes of objects \citep{Costa96}.  The SMC represents a
crucial metallicity and the detailed evolved star work proposed here
will provide strong constraints on dust production in the early
universe.

Figure~\ref{fig_smc_cmd_es} presents a preliminary look at the evolved
stars seen in the SAGE-SMC observations.  The different populations of
AGB stars are shown, including an unidentified group of AGB stars that
we call anomalous O-rich AGB stars (aO-AGB).  The IR properties of
these stars are similar to the O-AGB stars, but show slightly enhanced
emission from 3.6 to 24~\micron. These stars may be a dusty
sub-population of O-rich AGB stars, or they might be S-type AGB stars.
If the latter, then this will be the first time S-type AGB stars have
been identified photometrically in the Magellanic Clouds.  The evolved
stars are explored in detail in the companion SAGE-SMC evolved stars
paper \citep{Boyer11} and their mass loss properties
will be discussed in a follow-up paper \citep{Boyer11}.

\subsection{Star Formation}

The SMC offers a unique laboratory for studying tidally-driven star
formation.  It has been more profoundly affected by recent 
interactions than the LMC, both in terms of its overall morphology and
star formation history.  In addition, the SMC's low metallicity
influences its ISM properties in ways that 
impact on the physics of star formation.  The dust-to-gas ratio and
molecular gas content are lower in the SMC than in the LMC or MW, and
the UV radiation field is more pervasive.  These differences likely
lead to substantially altered star formation efficiencies, initial
mass functions, clustering properties of newly-formed stars, and size
and timescales for feedback and triggered star formation. Present day
SMC star formation is concentrated in the Bar, Wing, and Tail regions.
The Wing and Tail regions probe the most extreme star formation
conditions, more strongly affected by tidal interactions and at even
lower gas density than in the main body of the SMC.  Both IR SEDs of
unresolved Young Stellar Objects (YSOs) and the resolved behavior of
HII/star formation regions give insight into the SMC star formation.

\subsubsection{Young Stellar Objects}

\begin{figure*}[tbp]
\plotone{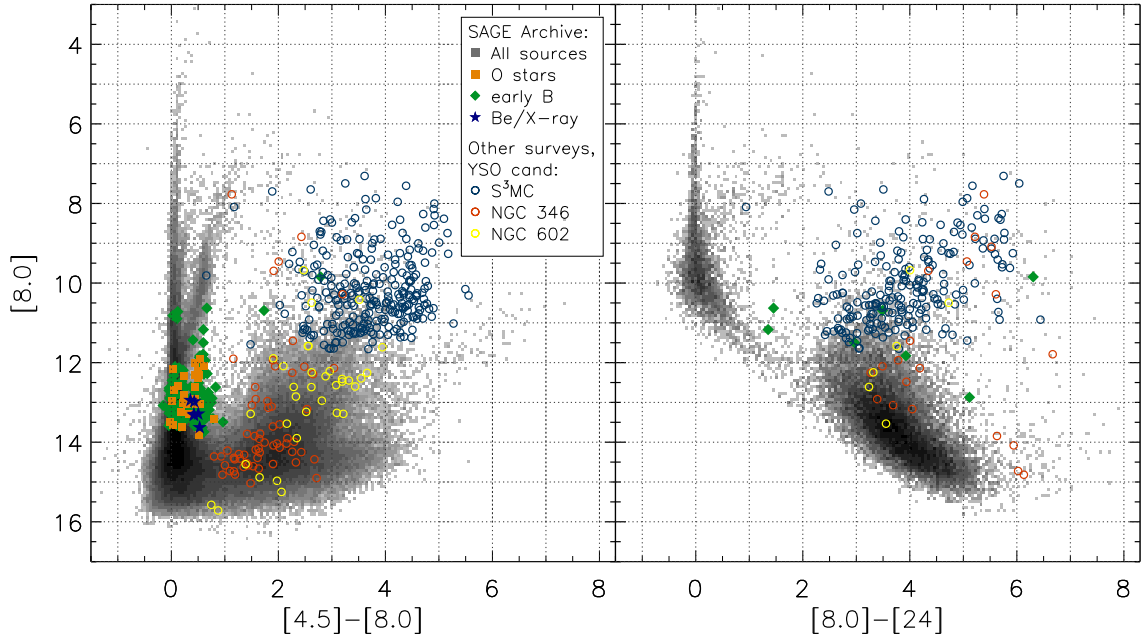}
\caption{The [8.0] vs. $[4.5]-[8.0]$ and $[8.0]-[24.0]$
color-magnitude diagrams for the SMC are shown created from the IRAC
SMP Archive and and MIPS 24~\micron\ epoch 1 full list.  The [4.5],
[8.0], \& [24.0] labels
give the Vega magnitude at those wavelengths.  Only sources which were
detected at $>$5$\sigma$ in all three bands 
are shown.  All sources are 
displayed as Hess diagrams ({\it grayscale}).  Overlaid are dusty
objects color coded according to the legend.  The O, early B, and
Be/X-ray sources are from \citet{Bonanos10}.  YSO
candidates from the S$^3$MC survey \citep{Bolatto07} and studies of
the individual star forming regions NGC~346 \citep{Simon07} and
NGC~602 \citep{Carlson11} are also shown. 
\label{fig_smc_cmd}}
\end{figure*}

SMC star formation historically has been
traced via HII regions, but with the advent of Hubble Space Telescope
and {\em Spitzer} 
systematic studies of YSOs are possible \citep{Chu05, Nota06}.
\citet{Bolatto07} found 280 high-mass YSOs by using the {\em Spitzer}
S$^3$MC pathfinder of the inner 3~$\sq\arcdeg$ of the SMC.  This number is
consistent with the few thousand high reliability YSOs discovered in
the SAGE-LMC data \citep{Whitney08, Gruendl09} when the different survey areas
are taken into account.  The full SMC YSO population is
being explored using the SAGE-SMC observations.
Fig.~\ref{fig_smc_cmd} presents two color magnitude diagrams (CMDs)
where the young objects are 
highlighted.  The IR properties of previously known massive stars have
been investigated by \citet{Bonanos10}.  YSO candidates from the
SAGE-SMC observations are explored in detail in the 
companion SAGE-SMC YSO paper (Sewi\'lo et al. 2011, in prep.).  The
embedded star formation in the SMC Tail \citep{Gordon09SMC} are
explored in detail in Chen et al. (2011, in prep.).

\subsubsection{HII/Star Formation Regions}

Prominent shell structures in both Magellanic Clouds suggest that
feedback may play an important role in shaping the pattern of star
formation \citep{Muller03, Zaritsky04, Hatzidimitriou05}. Proper
understanding of feedback timescales and mechanisms requires resolving
HII regions and CO clouds in the IR at the few parsec scale.  Of
particular interest is to study the star formation in the low density
environment of the Tail and how this affects the energy feedback
into the ISM.  Such an environment may be close to that expected at
the highest redshifts.  

\begin{figure}[tbp]
\epsscale{1.2}
\plotone{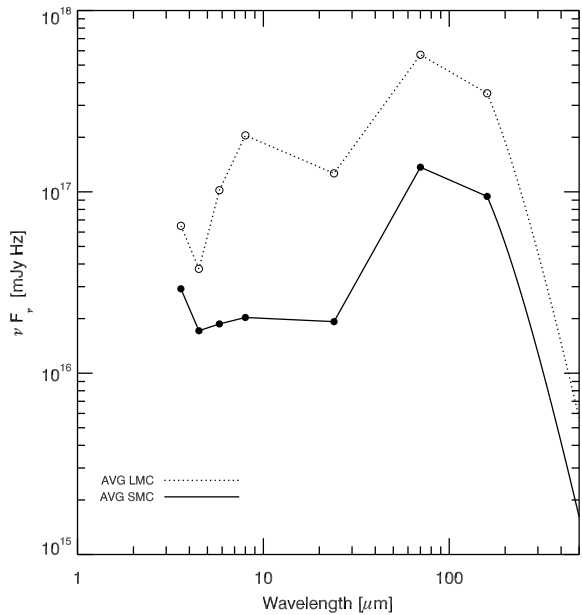}
\caption{The average IR SED of 151 SMC HII/star forming regions is shown
along with, for comparison, the average IR SED of 543 LMC regions.
The SMC HII regions have depressed mid-IR emission when compared to
the LMC HII regions, indicating significantly less aromatic feature
emission.
\label{fig_hii_sed}}
\end{figure}

The infrared SEDs of individual HII/star formation (SF) regions in the SMC
(and LMC) reveal the heating of the surrounding dust and the impact of
star formation on this dust.  In \citet{Lawton10}, the behavior of
different monochromatic obscured star formation rate (SFR) indicators was
compared to the total IR (TIR) SFR in the SMC and LMC using a small
sample of regions in each 
galaxy.  The 70~\micron\ band was found to be the best 
monochromatic SFR indicator based on the constancy of the
70~\micron/TIR flux ratio both within and between regions.  Unlike the
70~\micron/TIR ratio, the 8~\micron/TIR ratio varied significantly
from region to region and systematically between the LMC and SMC with
the SMC showing a much lower 8~\micron/TIR ratio.  This is illustrated
in Fig.~\ref{fig_hii_sed} where the average SEDs of SF regions in the
SMC and LMC are plotted.  The companion
SAGE-SMC star forming regions paper (Lawton et al. 2011, in prep.)
will explore the SEDs of SF regions in more detail including a complete
catalog of IR selected SF regions.

\section{Conclusions/Summary}

The motivation of the SAGE-SMC Sptizer Legacy was to study the
lifecycle of the dust in the SMC; from birth in the winds of evolved
stars, migration from the diffuse to dense ISM, and use as shielding
and fuel in the formation of stars.  The SMC provides a unique
laboratory for the study of the lifecycle of dust given its low
metallicity and relative proximity.  The SAGE-SMC {\em Spitzer} 
Legacy observed the full SMC from 3.6 to 
160~\micron\ using the IRAC and MIPS instruments.  The region surveyed
30~$\sq\arcdeg$ encompassing the SMC Body (Bar and Wing) and Tail
(high-density portion of the Magellanic Bridge).  The observations
were carefully reduced and high quality mosaics and point source
catalogs were produced.  These data products are available from the {\em Spitzer}
Science Center and IRSA Archive.  

Initial results from the SAGE-SMC
Legacy Team include finding that the gas-to-dust ratio in the SMC Tail
region matches that of the SMC Body \citep{Gordon09SMC}, probing the
detailed CO and dust structure of an SMC molecular cloud to confirm
that CO is significantly dissociated at cloud edges
\citep{Leroy09SMC}, studying the 
dust mass loss in the Galactic clusters NGC~362 \citep{Boyer09} and
47~Tuc \citep{Boyer10, McDonald11}, investigating the best monochromatic
obscured SFR indicator \citep{Lawton10}, studying the IR properties of
known massive stars \citep{Bonanos10}, and a detailed investigation of
the star formation in NGC~602 \citep{Carlson11}.
A set of companion papers to
this overview paper will investigate the interstellar dust (Bot et al. 2011, in
prep.), evolved stars \citep{Boyer11}, young stellar
objects (Sewi\'lo et al. 2011, in prep.), and HII/star formation regions
(Lawton et al. 2011, in prep.).  The launch of the {\em Herschel Space
Observatory} has enabled the infrared dataset on the SMC (and LMC) to be
extended in the submm through the HERITAGE Key Project
\citep{Meixner10}.  The future of Magellanic Clouds research in the
infrared is very bright.

\acknowledgements
This work is based on observations made
with the {\em Spitzer Space Telescope}, which is operated by the Jet
Propulsion Laboratory, California Institute of Technology under NASA
contract 1407. Support for this work was provided by NASA through
Contract Number \#1340964 issued by JPL/Caltech.

{\em Facility:} \facility{Spitzer (IRAC, MIPS)}

\appendix

\section{Corrections to S$^3$MC data}
\label{sec_s3mc_cor}

We combined the S$^3$MC and the SAGE-SMC observations to achieve
deeper maps in the regions of overlap.  In order to make this
combination, we had to process the MIPS 24 and 70~\micron\ S$^3$MC
observations to correct their background levels and remove residual 
instrumental effects.  These corrections to the S$^3$MC data were possible
once the SAGE-SMC observations were taken.  The SAGE-SMC data provide
a reference as they were taken with sufficient background measurements
(taken at the beginning and end of each scan leg) to allow for removal
of astronomical backgrounds to the SMC and residual instrumental
signatures.  The S$^3$MC MIPS 160~\micron\ observations did not
require any additional processing attesting to the excellent stability
of the MIPS 160~\micron\ data after standard data reduction steps.

\begin{figure*}[tbp]
\epsscale{1.1}
\plotone{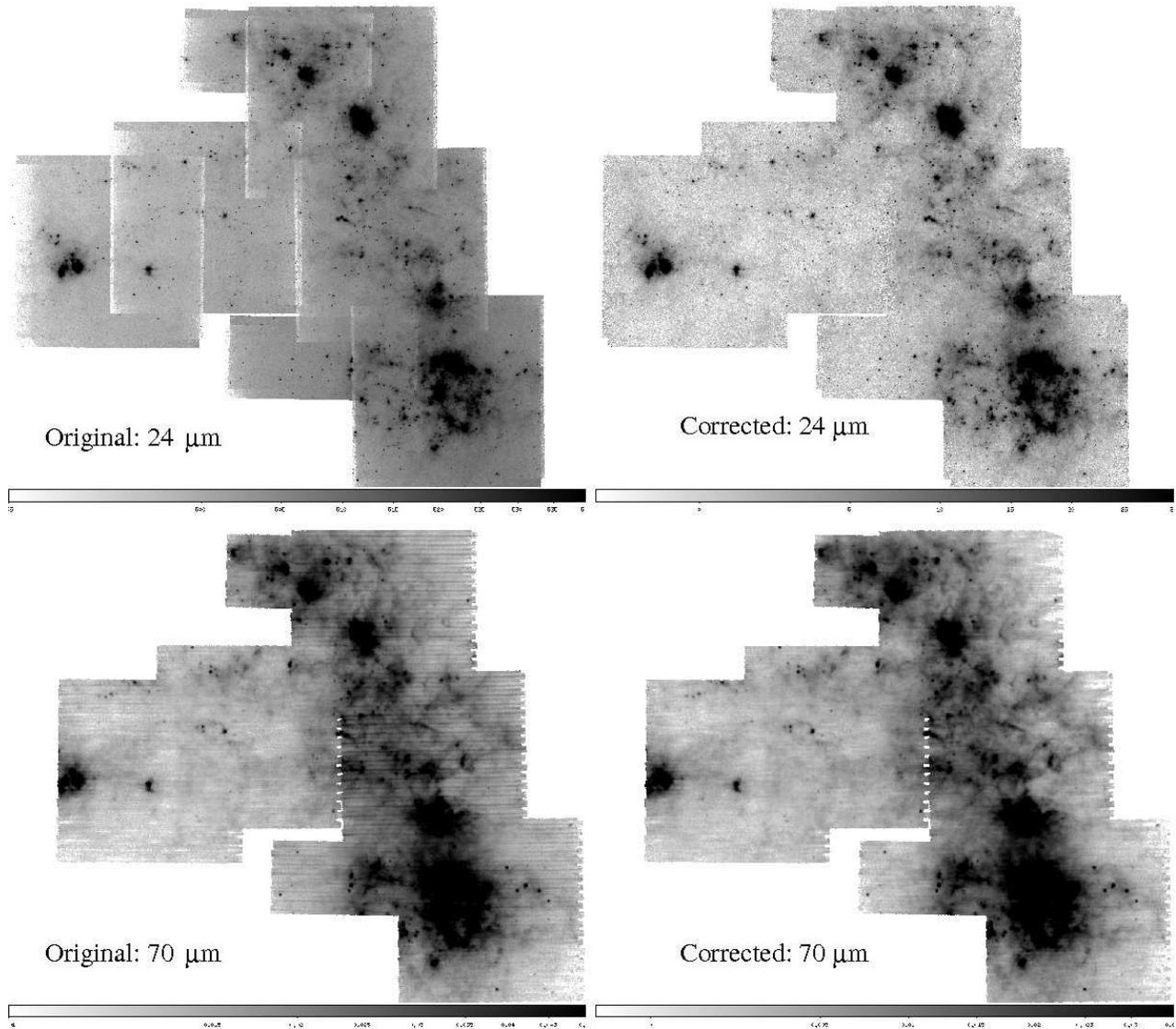}
\caption{The original and corrected S$^3$MC mosaics are shown for
24~\micron\ (top) and 70~\micron\ (bottom).  The images are displayed
with a square root stretch.  The corrections were done using the full
SAGE-SMC mosaics as a reference. 
\label{fig_s3mc_cor}}
\end{figure*}

At 24~\micron, the background is dominated by the
variable zodiacal light with a minor contribution from residual
instrumental variations in the first few images in a scan leg.  The
S$^3$MC observations were corrected by determining the difference
between the average level in each S$^3$MC image and that expected
using the SAGE-SMC epoch 1 mosaic (which already has zodiacal
light subtracted).  The temporal trend of these differences for each
scan leg was fit with a low order polynomial and the resulting fit
subtracted from the S$^3$MC data.  The correction in general flattened
the background removing a slight gradient from top to bottom as well
as instrumental signatures associated with the start of the scan
legs (see Fig.~\ref{fig_s3mc_cor}).  

At 70~\micron, the main issue with S$^3$MC data is seen as striping
in the maps (Fig.~\ref{fig_s3mc_cor}) that 
is caused by the baseline of different 70~\micron\ pixels drifting
with respect to the fast time constant calibration.  Since this effect
is already removed from the SAGE-SMC mosaic, we used it as a reference
for correcting the S$^3$MC data.  For each pixel, the difference
between the S$^3$MC measurement and that expected from the SAGE-SMC
epochs 1+2 mosaic was computed.  A low order polynomial was fit to the
temporal trend of the differences in each scan leg for each pixel and
the resulting fit subtracted from the S$^3$MC data for the appropriate
pixel.  The corrected S$^3$MC 70~\micron\ mosaic is shown in
Fig.~\ref{fig_s3mc_cor} and the scan direction streaking can be seen
to be significantly suppressed.

\end{document}